\documentclass {article}
\newtheorem{theorem}{Theorem}

\begin{document}

\author{Peter Kuchment\thanks{%
The work of both authors was partially supported by an NRC COBASE Grant. The
work of the first author was also supported in part by the NSF Grant DMS
9610444 and by a DEPSCoR Grant.} \\
Mathematics and Statistics Department\\
Wichita State University\\
Wichita, KS 67260-0033, USA \and Sergei Levendorski \\
Mathematics Department\\
Rostov State Academy of Economics\\
Rostov-on-Don, Russia}
\title{On absolute continuity of spectra of periodic elliptic operators}
\date{}
\maketitle

\begin{abstract}
The paper contains a brief description of a simplified version of A.
Sobolev's proof of absolute continuity of spectra of periodic
magnetic Schr\"{o}dinger operators. This approach is applicable to all
periodic elliptic operators known to be of interest for math physics
(including Maxwell), and in all these cases leads to the same model problem
of complex analysis. The full account of this approach will be provided
elsewhere.
\end{abstract}

\section{Introduction}

\footnotetext{\textit{1991 Mathematics Subject Classification}. Primary
35P99, Secondary 35J10
\par
\textit{Key words and phrases}. Periodic operator, elliptic operator,
absolutely continuous spectrum, Schr\"{o}dinger operator, magnetic and
electric potential}Elliptic differential operators with periodic
coefficients arise naturally in many areas of mathematical physics. One can
mention quantum solid state theory, where the main operators of interest are
the stationary Schr\"{o}dinger operator 
\begin{equation}
-\Delta +V(x)  \label{Schred}
\end{equation}
and the magnetic Schr\"{o}dinger operator 
\begin{equation}
(i\partial -A(x))^2+V(x).  \label{Magn}
\end{equation}
Here the scalar electric potential $V(x)$ and the vector magnetic potential $%
A(x)$ are assumed to be periodic with respect to a lattice in $\bf{R}^d$ 
\cite{AM}. Another example, which gained importance due to recent advances
in the photonic crystals theory \cite{JMW} is the periodic Maxwell operator 
$ \nabla \times \varepsilon (x)^{-1}\nabla \times $ 
defined on zero-divergence vector fields in $\bf{R}^3$. Here the periodic
function $\varepsilon (x)$ represents the electric permittivity of the
medium. Scalar counterparts of this operator (arising also in acoustics) are
the operators $-\nabla \cdot \varepsilon (x)^{-1}\nabla $ and $-\varepsilon
(x)^{-1}\Delta .$ Other examples are the anisotropic divergent type
operators $\sum \partial _ia_{ij}(x)\partial _j$, where the magnetic and
electric potential terms can also be included, and the periodic Dirac
operator (\cite{BS}, \cite{Da}, \cite{Mo}).

For all these operators, the structure of the spectrum is of major interest.
Consider a periodic Schr\"{o}dinger operator (\ref{Schred}) in $\bf{R}^d$.
Under mild conditions on the real potential the operator is self-adjoint 
\cite{RS}. It has been clear to physicists for a long time that the spectrum
of this operator in $L_2(\bf{R}^d)$ does not contain any eigenvalues. One
can easily prove absence of eigenvalues of finite multiplicity \cite{Ea},
however the proof of the general statement had to wait for a long time until
the celebrated Thomas' theorem \cite{T} (see also its improved version in 
\cite{RS}, \cite{Sh}). The paper \cite{T} contained the proof of a more general
statement: the spectrum of a selfadjoint periodic operator (\ref{Schred}) is
absolutely continuous. Absence of singular continuous spectrum for periodic
elliptic operators holds in a very general situation and is a rather
straightforward consequence of Floquet theory (\cite{K}, \cite{RS}, \cite{Sj}%
).

The question arises on whether absence of eigenvalues is shared by all
periodic elliptic differential operators. It is known, however \cite{K} that
this is not true in general for elliptic operators of order four. Still, the
common belief is that periodic elliptic operators of second order do not
possess any eigenvalues. The talks presented at this conference by M. Birman
and T. Suslina described some of such results. There are non-standard
situations of mesoscopic physics and photonic crystal theory where one can
encounter point spectrum of periodic second order differential problems on
graphs \cite{KK1}. This, however, does not influence our
belief in absolute continuity of spectra of such problems in $\bf{R}^d$.
Let us list the known results. The Thomas' theorem \cite{T} was extended to
a broader class of periodic Schr\"{o}dinger operators in the M. Reed and B.
Simon's book \cite{RS}. L. Danilov \cite{Da} proved absolute continuity of
the spectrum for the case of the Dirac operator with a periodic scalar
potential. The result for the magnetic Schr\"{o}dinger operator (\ref{Magn})
was obtained by R. Hempel and I. Herbst in \cite{HH} for the case of small
magnetic potentials. Finally, in the remarkable papers by M. Birman and T.
Suslina \cite{BS} and A. Sobolev \cite{So} the full strength statement about
the magnetic Schr\"{o}dinger operator was proven. The elegant algebraic
approach of the paper \cite{BS} works only in dimension two. The paper \cite
{BS} lead A. Sobolev to an ingenious proof in arbitrary dimension \cite{So}.
A generalization of the result of \cite{Da} was obtained in \cite{BS2}. A
recent very interesting preprint by A. Morame \cite{Mo} treats an
anisotropic divergent type operator in $2D$.

Our initial goal was to get good grasp of the proof presented in \cite{So}.
We found that the initial proof can be significantly simplified, which in
particular leads to improvements in the result and to the possibility of
broad generalizations. Our aim is mostly methodological: to provide a
simplified and unified approach that treats in an uniform way all the
operators mentioned above and, we hope, clarifies the situation. It also
leads to some improvements in the result of \cite{So} (weaker assumptions on
the potential, a stronger estimate from below, and validity in
non-selfadjoint case). The interesting observation is that in all situations
known to the authors the same model $\overline{\partial }$ problem arises,
good understanding of which would lead to progress in all cases that are not
treated yet (Maxwell, divergent, etc.). Most of the ideas of the proof were
already contained in a more obscured form in papers \cite{BS} and \cite{So}.
This paper contains only a brief description of the results. The complete
account will be given elsewhere.

\section{The magnetic Schr\"{o}dinger operator}

Our main object of study is the magnetic Schr\"{o}dinger operator in $\bf{R}%
^d$%
\[
H=(D+A(x))^2+V(x)
\]
with a scalar electric potential $V(x)$ and a vector magnetic potential $A(x)
$. We will assume for simplicity of presentation that both potentials are
periodic with respect to the integer lattice $\bf{Z}^d$; the case of
general lattices does not require any significant changes in the proofs.
According to Thomas \cite{T} (see also \cite{RS}, \cite{K}), absence of
eigenvalues will be proven if one is able to show existence of a \textit{%
quasimomentum} $k\in \bf{C}^d$ such that the operator 
\[
H(k)=(D+k+A(x))^2+V
\]
has zero kernel on the torus $\bf{T}^d=\bf{R}^d/\bf{Z}^d$. Our key
statement is the following theorem:

\begin{theorem}
\label{Estim}Let $A\in \left[ H^s(\bf{T}^d)\right] ^d$ for some $s>3d/2-1$.
Then there exist constants $C>0$, $\beta >0$ and vector $e\in \bf{R}^d$,
such that for sufficiently large $\rho \in \bf{R}$ and any $u\in H^2(\bf{T}%
^d)$%
\[
\left| \left| (D+k+A(x))^2u\right| \right| _{L_2}\geq C(\left| \rho \right|
\left| \left| u\right| \right| _{L_2}+\left| \left| u\right| \right| _{H^1}),
\]
where $k=(\beta +i\rho )e\in \bf{C}^d$.
\end{theorem}

An immediate consequence of this estimate is the following statement.

\begin{theorem}
Assume that $A$ is like in the previous theorem and the electric potential $V
$ is such that 
\begin{equation}
\left| \left| Vu\right| \right| _{L_2(\mathbf{{T}^d})}\leq C_\varepsilon
\left| \left| u\right| \right| _{L_2(\mathbf{{T}^d})}+\varepsilon \left|
\left| u\right| \right| _{H^1(\mathbf{{T}^d})}  \label{el_pot}
\end{equation}
for arbitrary $\varepsilon >0$ and $u\in H^1$ (for instance, $\left| \left|
Vu\right| \right| _{L_2(\mathbf{{T}^d})}\leq C\left| \left| u\right| \right|
_{H^\alpha (\mathbf{{T}^d})}$ for some $\alpha <1$). Then the periodic
magnetic Schr\"{o}dinger operator $H=(D+A(x))^2+V(x)$ has no point spectrum
in $L_2(\mathbf{{R}^d)}$.
\end{theorem}

The condition (\ref{el_pot}) can be verified for different classes of
potentials, as it was done in the previously mentioned works on this 
subject (the details will be added in the complete version). 

In the case of real potentials the operator is self-adjoint, and one obtains
the following result.

\begin{theorem}
The spectrum of the periodic magnetic Schr\"{o}dinger operator $%
H=(D+A(x))^2+V(x)$ in $L_2(\mathbf{{R}^d)}$ is absolutely continuous.
\end{theorem}

\section{The scheme of the proof}

One can consider the case of zero electric potential, since a standard
procedure enables one to include the electric potential (see for instance 
\cite{K}, \cite{RS}, \cite{So}).

Let $H(k)=(D+k+A(x))^2$, and $\Lambda _{\,\rho }$ be the operator that
multiplies the $m$th Fourier coefficient of a periodic function by $(\rho
^2+m^2)^{1/2}$. These are operators on the torus $\bf{T}^d$. Let 
\[
k=2\pi (i\rho +\beta )e\in \bf{C}^d,
\]
where $\beta \in \bf{R}$ is fixed, $\rho \in \bf{R}$ is arbitrarily large,
and $e\in \bf{R}^d$. We introduce the principal symbol of the operator $H(k)
$ as 
\[
H_0(k,m)=(2\pi m+k)^2=4\pi ^2\left[ (m+\beta e)^2-\rho ^2+2i\rho e\cdot
(m+\beta e)\right] .
\]
Notice that we include some lower order differential terms with parameter $k$
into the principal symbol, which is rather standard when working with
pseudodifferential operators with parameters. The zero set of this symbol is 
\[
Z_\rho =\left\{ m|\;(m+\beta e)^2=\rho ^2,\,e\cdot (m+\beta e)=0\right\} .
\]
We choose a suitable finite multiplicity covering of the dual space $\bf{R}%
_{\,\xi }^d$ by sets $U_{\rho j}$ of diameter $\rho ^{\,\delta }$ for an
appropriately chosen $\delta \in (0,1)$. The goal is to produce a set of
local approximate inverse operators $R_{\rho ,j}$ such that on functions
whose Fourier series are supported in $U_{\rho j}$ the following properties
are satisfied: $\left| \left| R_{\rho ,j}\right| \right| \leq C$, $R_{\rho
,j}H(k)\Lambda _{\,\rho }^{-1}=I+T_{\rho ,j}$, and $\left| \left| T_{\rho
,j}\right| \right| \rightarrow 0$ uniformly with respect to $j$ when $\rho
\rightarrow \infty $. Then a suitable partition of unity in the dual space
finishes the job. Namely, an operator 
\[
R_{\,\rho }=\sum_j\phi _j(D)R_{\rho ,j}\psi _j(D)
\]
is constructed in such a way that 
\[
R_{\,\rho }H(k)\Lambda _{\,\rho }^{-1}=I+T_{\,\rho },
\]
where 
\begin{equation}
\lim_{\rho \rightarrow \infty }\left| \left| T_{\,\rho }\right| \right| =0.
\label{zero}
\end{equation}
Here $\phi _j(D)$ and $\psi _j(D)$ are operators that multiply Fourier
coefficients of a function by cut-off functions $\phi _j$ and $\psi _j$.
Existence of such an operator $R_{\,\rho }$ proves the Theorem \ref{Estim}.

Now we see that the main task is to construct the local approximate inverse
operators $R_{\rho j}$. The situation looks differently away from the set $%
Z_{\,\rho }$ and closely to it. Namely, when the distance from $U_{\rho j}$
to $Z_{\,\rho }$ is more than $\rho ^{\,\delta }$, the principal part will
dominate the magnetic one, which analogously to the Thomas' case leads to
invertibility. However, close to $Z_{\,\rho }$ the magnetic potential part
becomes of comparable strength with the principal part. In each of these
open sets the operator $H(k)\Lambda _{\,\rho }^{-1}$ can be reduced to a
model first order differential operator by linearizing its symbol at one
point. This model operator happens to be of a generalized Cauchy-Riemann
type 
\begin{equation}
\frac \partial {\partial \overline{z}}+g(z)  \label{model}
\end{equation}
on a complex plane, where the plane arises as a rational plane in $\bf{R}^d$
spanned by two integer vectors $l$ and $n$, and the function $g(z)$ is
periodic. The goal is to invert such an operator on the torus with
controlled norm of the inverse. A. Sobolev invented a trick that does this 
\cite{So}. Namely, one can gauge away most of the magnetic potential $A$,
leaving only a small part of it, which allows to construct a controllable
inverse operator. The construction of the gauge transform amounts to solving
the Cauchy-Riemann equation $\overline{\partial }f=g(z)f$ for an invertible
periodic function $f$ on the plane. In general there are obstructions (some
Fourier coefficients of $g$ must vanish), but a clever choice of an $l-n$
plane guarantees that these obstructions are moved into a tail of the
Fourier series of the potential. This allows one to gauge away most of the
potential and to construct an approximate inverse with controllable norm.

\section{Concluding remarks and acknowledgments}

This scheme works uniformly in all cases of interest. It enables one to
obtain in the same way (and with the same model local operator) the results
of \cite{BS}, \cite{BS2}, \cite{Da}, \cite{So}, and \cite{T}. The scheme
also works in all cases where complete results are not yet known: the
Maxwell operator, Dirac operator with a general matrix potential, and
anisotropic divergent type operators. In all these three situations one
discovers that the local model operator on $Z_{\,\rho }$ (\ref{model}) has a
matrix ''potential'' $g$. One ends up with the following question: given a
periodic matrix function $g(z)$ on the complex plane one needs to find an
invertible periodic matrix function $f(z)$ such that $\overline{\partial }%
f=g(z)f$. So, we have to deal with a non-commutative version of the model
problem that was discussed above. This problem has been studied in complex
analysis \cite{Ma}. The function $g$ defines in some natural way a
holomorphic vector bundle, and a required function $f$ exists if and only if
this bundle is trivial. What one needs to know for the spectral problem that
we discuss is whether there is any ``non-commutative'' analog of Sobolev's
trick that moves the obstruction into a long tail of the Fourier expansion
and gauges away the rest of the series.

The authors express their gratitude to Professors M. Birman, A.
Sobolev, T. Suslina, A. Tumanov, R. Novikov, A. Pankov, and V. Palamodov for information
and useful discussions. This research was partly sponsored by the NRC
through a COBASE Grant, NSF through the Grant DMS 9610444, and by the
Department of Army, Army Research Office, through a DEPSCoR Grant. The
authors thank the NRC, NSF, and the ARO for this support. The content of
this paper does not necessarily reflect the position or the policy of the
federal government, and no official endorsement should be inferred.

\end{document}